\documentclass{llncs}
\usepackage{graphicx}
\usepackage{subcaption}
\begin{document}

\title{PerspectivesX: A Proposed Tool to Scaffold Collaborative Learning Activities within MOOCs}
\titlerunning{PerspectivesX}  
%
\author{Aneesha Bakharia}
\authorrunning{Aneesha Bakharia et al.} 
%
\tocauthor{Aneesha Bakharia}
\institute{UQx, ITaLI, The University of Queensland, St Lucia, Brisbane, Australia\\
\email{a.bakharia@uq.edu.au}}

\maketitle              

\begin{abstract}
In this work-in-progress paper, we introduce the PerspectivesX tool which aims to scaffold collaborative learning activities within MOOCs. The PerspectivesX tool has been designed to promote learner knowledge construction and curation for a range of multi-perspective elaboration techniques (e.g., SWOT analysis and Six Thinking Hats). The PerspectivesX tool is designed to store learner submissions in a searchable knowledge base which is able to be persisted across course re-runs and promotes the use of natural language processing techniques to allow course moderators to provide scalable feedback. In this paper we outline the design principles that structured collaborative learning tools need to adhere to, design a prototype tool (PerspectivesX) and evaluate whether MOOC platform extension frameworks are able to support the implementation of the tool.   
\end{abstract}
\section{Introduction}
The tool predominantly used in MOOCs to foster collaborative learning is the discussion forum. Research has shown that learners that actively contribute to the course forum, are more likely to complete the course and achieve higher grades \cite{corrin2017using}.  A high percentage of learners however, don't engage in a course discussion forum with recent estimates of forum participation being between 5-10\% of participants \cite{barriertoengagement}. There currently exists a wide gap between the unstructured collaborative nature of forums and other MOOC instructional content (i.e., videos, quizzes and social polls). Tools that are able to scaffold collaborative learning activities are required.

Computer Supported Collaborative Learning (CSCL) and the ideas behind providing scripted learning activities has a long and rich research history. Unfortunately, the theoretical underpinnings and practical manifestations of CSCL have all but been forgotten in the era of the MOOC. In this paper, the PerspectivesX tool is introduced. The PerspectivesX tool implements concepts from CSCL scripting; and the Knowledge Community and Inquiry model (KCI) \cite{Slotta2013}. KCI uses Web 2.0 tools to add a layer of collective knowledge building to scripted learning activities. 
\section{The PerspectiveX Tool}
The PerspectivesX tool is able to scaffold a range of multi-perspective elaboration activities. The tool is designed to promote active participation from learners that are either not participating in a discussion forum or that are passive forum participants (i.e., only reading forum posts). PerspectiveX encourages learners to make a contribution and also makes it easy for learners to explore, review and curate other learners submissions. In a PerspectivesX activity, learners must think about a problem from an assigned or selected perspective and actively contribute their ideas to a knowledge base that is available to all course participants. Instructors can enable an optional curation layer that requires learners to collate ideas from fellow learners in order to complete the remaining perspectives of the activity. Curation is an important feature of the tool. Curation is a 21st century digital literacy that is able to facilitate the development of learner search and evaluation strategies as well as promote critical thinking, problem solving, and participation in networked conversations \cite{oconnell2012}. 

Example activities that the PerspectiveX tool is able to scaffold includes learner submissions for design projects (i.e., knowledge construction), reflective journal entries (i.e., critical thinking) and multi-perspective elaboration activities (i.e., idea generation). The suggested approach will be able to support a range of idea generation and multi-perspective activities such Strengths, Weakness, Opportunities and Threats (SWOT) analysis, Six Thinking Hats \cite{de1999six}, Fishbowl \cite{miller2008fishbowl} and SCAMPER \cite{eberle1996scamper}.

\section{Design Principles}
The design principles that underpin the PerspectivesX tool are outlined below: 
\begin{itemize}
\item \textbf{Support the design of structured knowledge construction, critical thinking and multi-perspective elaboration activities}
\hfill \\
Instructors should be able to design activities that are able to collate structured responses/submissions from learners. The types of responses required by learners should be flexible and allow learners to submit multiple free text responses, media artifacts (e.g., images, infographics, slides, videos, etc) and links to external resources (e.g., website links). Within multi-perspective activities the instructor should be able to design activities that allow the learner to select a perspective or be randomly assigned to perspective.
\item \textbf{Support opt-in and anonymous learner knowledge sharing}
\hfill \\
Learners should not be forced to share their submissions with other course participants. Between 5-10\% of learners are active discussion forum participants in a MOOC while a larger percentage of learners read forum posts (i.e. passive participation). Many learners may not feel confident making their submissions available to other learners in a non-anonymous environment. Submission should be mandatory in order to receive a participation grade but the learner should be able to opt-out of sharing or choose to be anonymous.  
\item \textbf{Support instructor moderation}
\hfill \\
Course moderators need the ability to review and curate useful learner contributions. Curated content will help learners to focus their attention on relevant and important submissions \cite{boyd2010streams} from other learners. The learner should be able to view moderator highlighted content in an accessible and intuitive manner. This will give moderators the ability to use learner submitted work as a starting point to trigger active participation in a discussion forum.  
\item \textbf{Support learner curation} 
\hfill \\
The scripted collaborative activity should allow for the inclusion of a learner curation sub-activity. As an illustrative example, the collaborative activity might require the learner to submit a single section of a SWOT activity (e.g., strengths) and then at a later stage, curate content from other course participants for the other sections (e.g., weaknesses, opportunities and threats).
\item \textbf{Support temporal independence}
\hfill \\
Both paced and self-paced MOOCs should be able to include scaffolded collaborative learning activities.  Learners should be able to contribute to the activity at any time as well as review and curate the submissions of other learners in a time independent manner. This is particularly important for self-paced MOOCs where learners are able to commence a course at any time and as a result would engage in collaborative learning activities at different times. Discussion forums within self-paced MOOCs are also less active, giving learners limited opportunities to either actively or passively participate in collaborative learning activities. 
\item \textbf{Support knowledge base growth across course re-runs}
\hfill \\
Learner contributions should collectively form a knowledge base which becomes available across course re-runs offered in a variety of delivery modes (i.e., paced and self-paced). Initial course runs often have a higher number of enrolled learners and more discussion forum activity as a result. Each MOOC re-run, begins with a refreshed discussion forum which results in community knowledge between courses being lost. Retaining student contributions will facilitate knowledge growth but also poses information retrieval problems. The interface used to display learner contributions will need to therefore include intuitive navigation, free text and tag based (i.e., folksonomy) search functionality.   
\item \textbf{Facilitate the delivery of customised scalable feedback}
\hfill \\
While various Natural Language Processing (NLP) and Deep Learning algorithms exist, the ability to accurately grade and provide feedback for free text student submissions within MOOCs has not been realised. There are however techniques that can be used scale feedback provided by instructors, moderators and tutors. These techniques rely on the similarity between learner submissions and are able to cluster similar learner responses together. Topic modeling using the Latent Dirichlet Allocation (LDA) \cite{blei2003latent} algorithm is a promising document clustering technique that can be used to find common topics in student submissions. Instructors, moderators and tutors can then view a summary of the topics that exist in learner submissions and provide feedback. Various implementations of using clustering to provide feedback at scale have been discussed by \cite{mohler2009text}. The topic modeling summary provides an additional way for learners to gain an overview of other student submissions and navigate the community constructed knowledge base. 
\end{itemize}
\section{Design Prototype}
In this section, screen mockups for a prototype that adheres to all of design principles listed in the previous section are presented. Most tools that support pedagogical scripting of CSCL employ a visual flowchart metaphor \cite{faucon2017demo}. The flowchart metaphor allows the designer to sequence key stages in the activity and specify whether an individual or group will contribute to the activity. The flowchart metaphor provides a high level overview of the activity, but the instructor is still required to configure each stage of the activity. We take a declarative approach to the configuration of the activity, which both simplifies and reduces the steps required to use the tool. The declarative approach is encapsulated in a simple user interface that allows the instructor to configure the activity.

The activity creation interface (see Figure 1), allows instructors to choose a template and specify the activity configuration settings. The instructor can specify how learners contribute to the perspectives in an activity (i.e., the learner contributions section). Options are provided for the instructor to allow learners to choose a perspective, contribute to all perspectives, or be randomly assigned a perspective. The instructor is able to enable a curation stage and configure the knowledge base.
\begin{figure*}
\includegraphics[width=4in]{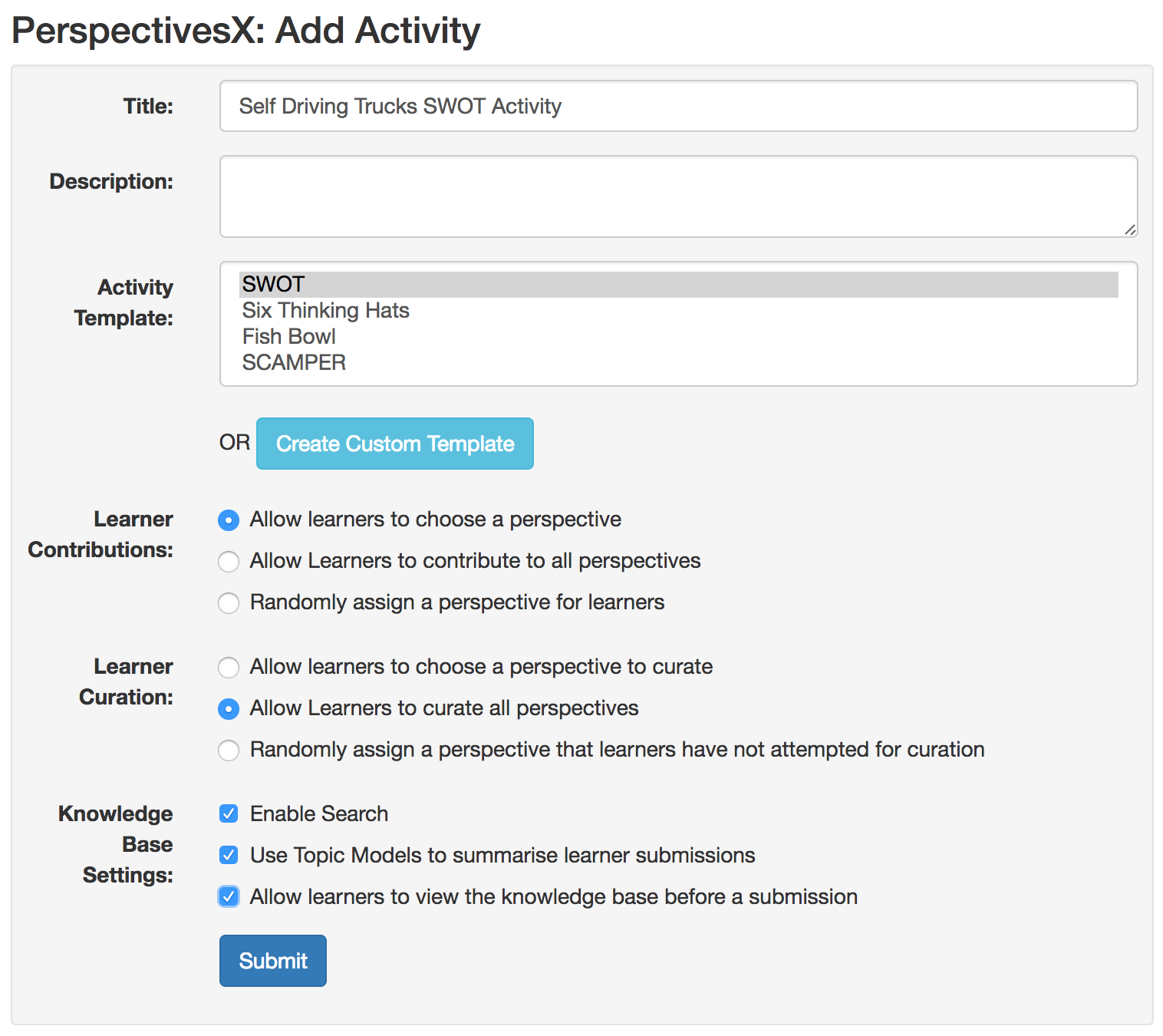}
\caption{The instructor multi-perspective activity creation interface.}
\end{figure*}

Central to the design of the PerspectivesX tool, is a structured template that instructors are able to create. It is envisaged that the tool will include standard templates for common activities such as Six Thinking Hats \cite{de1999six}, SCAMPER \cite{eberle1996scamper} and Fish Bowl \cite{miller2008fishbowl}. Instructors will also be able to create custom templates. As an example, a template can be created for a SWOT activity using a multi-perspective fieldset to include each text contribution field that is required (i.e., Strengths, Weaknesses, Opportunities and Threats). The interface to create a template is shown in Figure 2a.


An example learner submission user interface is displayed in Figure 2b. The fields that a learner is required to complete is dependent upon the settings the instructor has selected. In Figure 3, the learner has to select a perspective, enter their contribution and decide whether their contribution will be shared with other students. A knowledge base is displayed after a learner submits their perspective. The learner is able to see their contribution as well as view other student contributions that have similar or opposing views. Content curated by a moderator will be included.    

\begin{figure}
\begin{subfigure}{.5\textwidth}
  \centering
  \includegraphics[width=.8\linewidth]{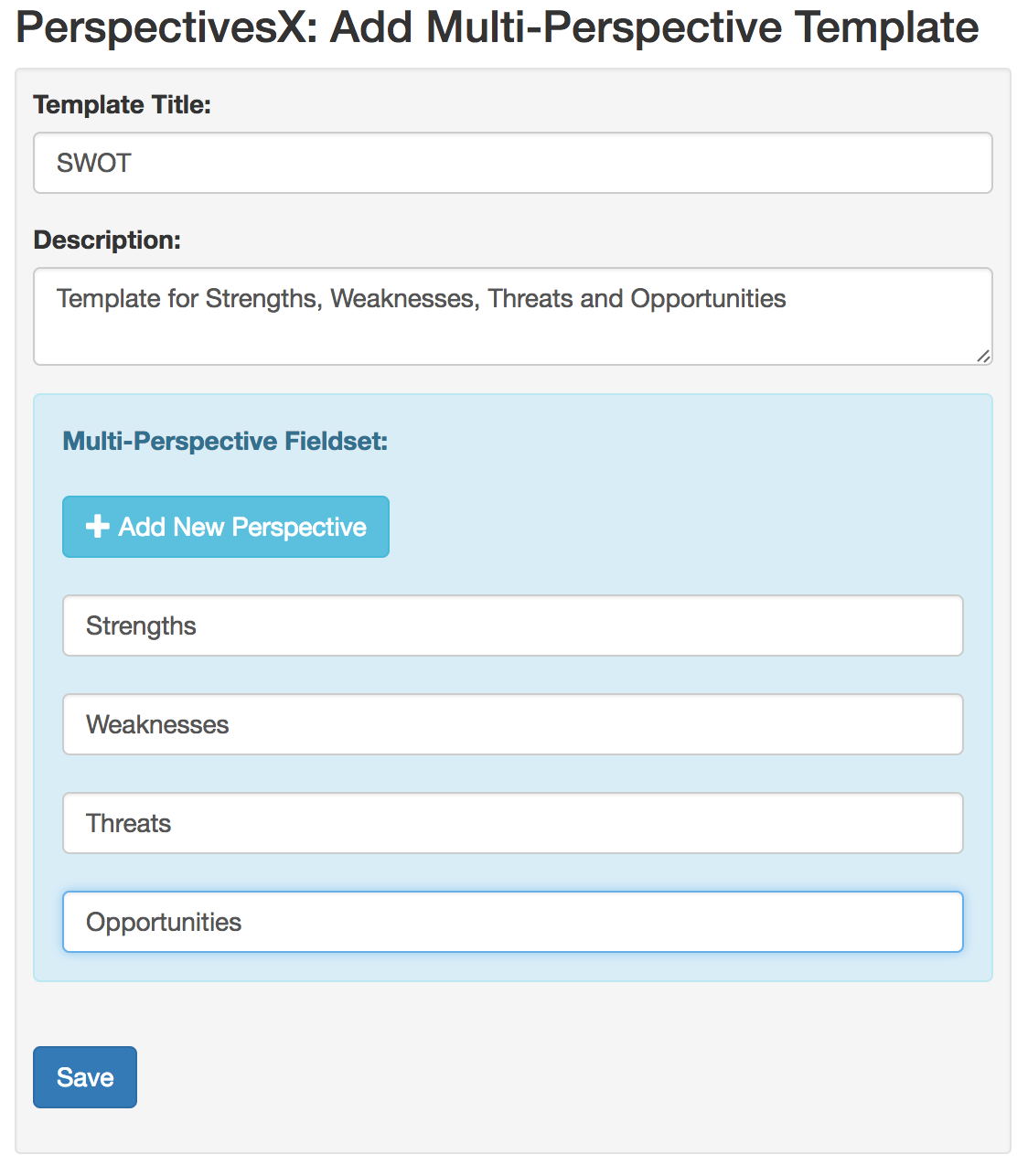}
  \caption{The multi-perspective template creation interface.}
  \label{fig:sfig1}
\end{subfigure}%
\begin{subfigure}{.5\textwidth}
  \centering
  \includegraphics[width=.8\linewidth]{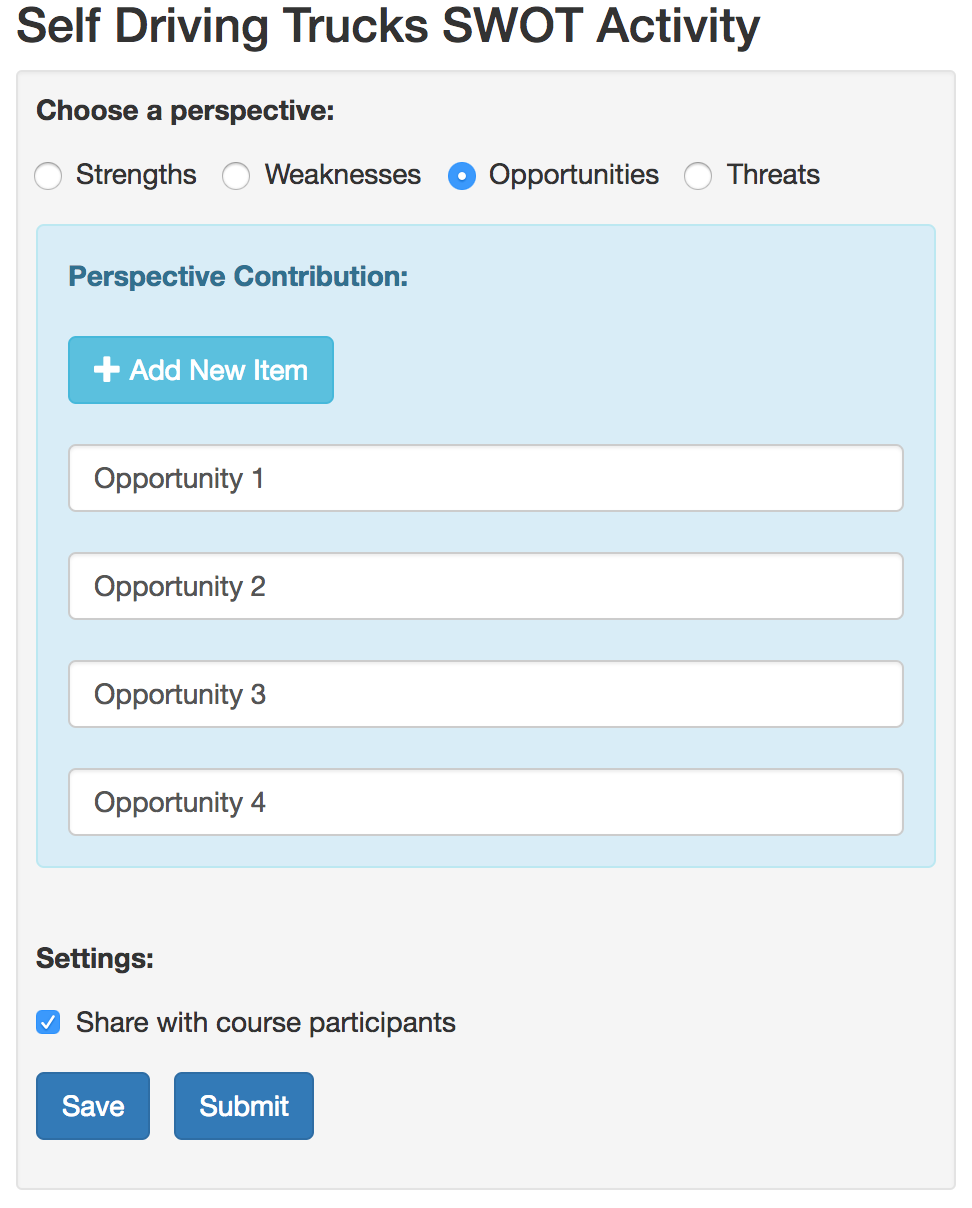}
  \caption{The learner contribution interface.}
  \label{fig:sfig2}
\end{subfigure}
\caption{Additional screen designs from the PerspectivesX tool.}
\label{fig:fig}
\end{figure}
\section{Implementation Considerations: LTI Tool vs XBlock}
The PerspectivesX tool can either be implemented using the Learning Tools Interoperability (LTI) specification or as an XBlock for the Open edX platform. LTI tools can be built in any programming language, have their own user interface and are able to run on their own server. LTI tools are also able to integrate with a range of Learning Management Systems that implement the LTI specification. XBlocks are extensions for the Open edX platform, must be built in the Python programming language and adhere to the Open edX user interface standard. Both the LTI and XBlock implementation options are comparable in terms of creating a user interface for the instructor and learner. As LTI's have the flexibility of being installed on a separate server, key features for the knowledge base will be easier to implement and scale. These features include the persistence of knowledge base data across course-runs and content indexing for search. Implementing PerspectivesX as an LTI would provide more flexibility to readily integrate with advance NLP and Deep Learning algorithms. 

\section{Conclusion and Future Directions}
%
The PerspectivesX tool will be developed as an open source LTI tool from the design mockups proposed in this paper. Future research will focus on the evaluation of the PerspectivesX tool and extending the design principles to support synchronous collaborative activities. 

\bibliographystyle{splncs03} 
\bibliography{bibfile}

\end{document}